\documentclass[aps, superscriptaddress,amsmath,amssymb,floatfix, notitlepage, bm,braket, prb, twocolumn, longbibliography]{revtex4}
\usepackage{times}
\usepackage{graphics}
\usepackage{graphicx}
\usepackage{color}
\usepackage{bm}
\usepackage{braket}
\usepackage{mathtools}
\usepackage{mathcomp}
\usepackage[caption=false,justification=raggedright,position=top,singlelinecheck=off]{subfig}
\usepackage{soul}
\usepackage{tabularx}
\usepackage{calrsfs}
\usepackage[vcentermath]{youngtab}
\usepackage[mathscr]{euscript}

\newcommand{\enbe}{\begin{equation}}
\newcommand{\enee}{\end{equation}}
\newcommand{\enba}{\begin{align}}
\newcommand{\enea}{\end{align}}

\begin{document}

\title{Skyrmions in twisted van der Waals magnets}
\author{Muhammad Akram}
\author{Onur Erten}
\affiliation{Department of Physics, Arizona State University, Box 871504, Tempe 85287-1504, AZ, USA}

\begin{abstract}
Magnetic skyrmions in two-dimensional (2D) chiral magnets are often stabilized by a combination of Dzyaloshinskii-Moriya interaction and external magnetic field. Here, we show that skyrmions can also be stabilized in twisted moir\'e superlattices with Dzyaloshinskii-Moriya interaction in the absence of an external magnetic field. Our setup consists of a 2D ferromagnetic layer twisted on top of an antiferromagnetic substrate. The coupling between the ferromagnetic layer and the substrate generates an effective alternating exchange field. We find a large region of skyrmion crystal phase when the length scales of the moir\'e periodicity and skyrmions are compatible. Unlike chiral magnets under magnetic field, skyrmions in moir\'e superlattices show enhanced stability for the easy-axis (Ising) anisotropy which can be essential to realize skyrmions since most van der Waals magnets possess easy-axis anisotropy.
\end{abstract}

\maketitle

\section{Introduction}
The discovery of ferromagnetism in two-dimensional (2D) monolayer CrI\textsubscript{3} and other 2D van der Waals (vdW) materials opened a new window for exploring low dimensional magnetism and its applications in spintronics\cite{gibertini2019magnetic, gong2019two, burch2018magnetism, huang2017layer, gong2017discovery, deng2018gate, bonilla2018strong, o2018room, soriano2019interplay}. The properties of 2D materials can be controlled by external parameters\cite{cristoloveanu2019concept,ye2012superconducting,cao2018unconventional,huang2018electrical,song2019switching} and are highly sensitive to stacking and twisting between the layers\cite{ponomarenko2013cloning,dean2013hofstadter,gorbachev2014detecting,cao2018correlated,cao2018unconventional}. In particular, with the discovery of superconductivity in twisted bilayer graphene\cite{cao2018correlated,cao2018unconventional}, there has been tremendous progress on exploring moir\'e superlattices both experimentally and theoretically~\cite{zhang2017interlayer,moon2014electronic,chen2019evidence,chen2019signatures,wallbank2015moire,ni2015plasmons}. In terms of magnetism, stacking order and twisting can significantly alter the interlayer exchange as the exchange is highly sensitive to atomic registries\cite{Chen_Science2019, Sivadas_NanoLett2018, Jiang_PRB2019, du2016weak,klein2018probing, tong2018skyrmions}.  

Magnetic skyrmions~\cite{roessler2006spontaneous} are nanoscale  vortex-like spin textures that were first observed in non-cetrosymmetic bulk magnetic materials such as MnSi~\cite{muhlbauer2009skyrmion,yu2010real}, (FeCo)Si~\cite{munzer2010skyrmion} and FeGe~\cite{yu2011near}. Skyrmions are topological defects, and they exhibit novel transport phenomena such as topological Hall effect and topological Nerst effect~\cite{hamamoto2015quantized, shiomi2013topological}. In recent years skyrmions received ample attention due to their potential for  spintronics applications and memory storage devices~\cite{fert2013skyrmions}. In most cases, skyrmions are stabilized by interplay of Dzyaloshinskii-Moriya (DM) interaction and external magnetic field\cite{muhlbauer2009skyrmion,yu2010real,banerjee2014enhanced}. In this article, we explore the possibility of stabilizing magnetic skyrmions in the absence of an external magnetic field in moir\'e superlattices. We consider a ferromagnetic (FM) monolayer on an antiferromagnetic (AFM) substrate with N\'eel order. Twisting the FM layer by an angle $\theta$ produces moir\'e patterns as shown in Fig. 1(a), (c). Ferromagnetic coupling between the substrate and the FM monolayer leads to an alternating exchange field for the moir\'e superlattice as shown in Fig. 1(b). Our setup is motivated from Ref. \onlinecite{tong2018skyrmions}. However unlike Ref. \onlinecite{tong2018skyrmions} which includes dipole-dipole interaction to stabilize magnetic skyrmions, we consider DM interaction which is the primary interaction for magnetic skyrmions in chiral magnets\cite{muhlbauer2009skyrmion, yu2010real, johnson1996magnetic, dubowik1996shape, Banerjee_NatPhys2013}.
\begin{figure}[t]
\includegraphics[width=\columnwidth]{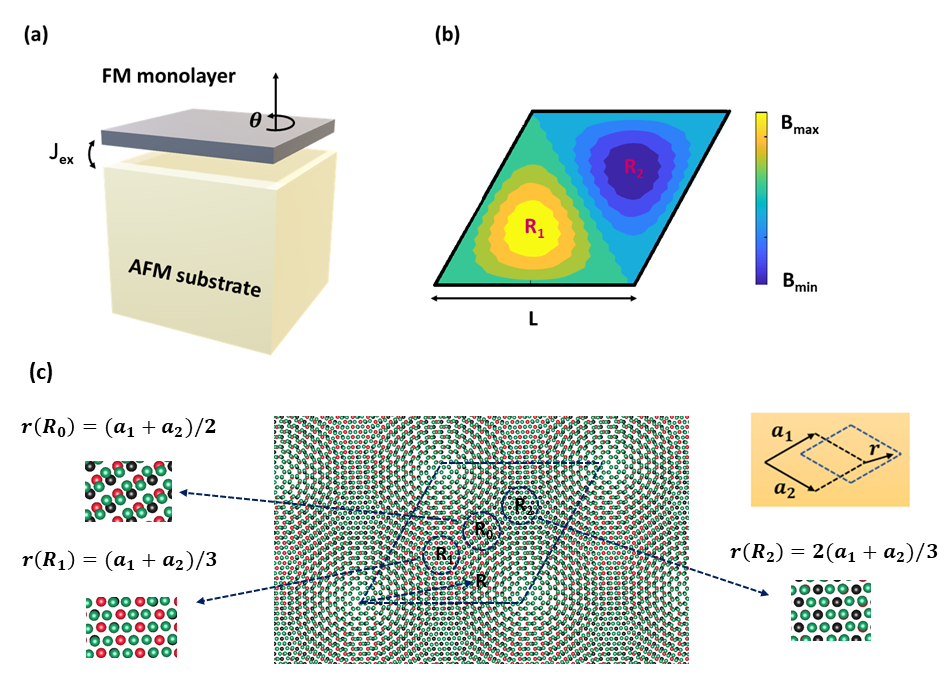}
\caption{(a) Schematic of our setup of a FM monolayer twisted on top of an AFM substrate with N\'eel order. We consider honeycomb lattice for both the monolayer and the substrate. (b) Exchange field due to interlayer coupling. (c) Moir\'e pattern due to small twisting angle between FM monolayer (green) and top most layer of AFM substrate (red and black). Three local atomic registries $\textbf{R}_0$, $\textbf{R}_1$ and $\textbf{R}_2$ are zoomed which matches the atomic registries of different interlayer translations.}
\label{Layer+substrate}
\end{figure}

Our main results are summarized in Fig.~\ref{L_vs_B_{Max}} and~\ref{A_vs_B_{Max}}. We show that (i) skyrmion crystal (SkX) is stabilized as a function of exchange coupling between the layers ($J_{\rm ex}$) and moir\'e periodicity. (ii) Even though SkX can be stabilized for a wide range of twisting angle, we find the optimal moir\'e periodicity to be about $L=9L_D$, where $L_D=(J/D)a$ is the intrinsic length scale for skyrmions. (iii) Unlike chiral magnets under magnetic field, we find an extended region of SkX for easy-axis anisotropy. (iv) We show that a large fraction of the topological charge $q=\frac{1}{4\pi}\int d^2r \widehat{\textbf{m}}.(\partial_{x}\widehat{\textbf{m}}\times\partial_{y}\widehat{\textbf{m}})$ of the magnetic skyrmions is concentrated at the edges and splits into three parts for large moir\'e periodicity and large easy axis anisotropy. This effect arises due to the anisotropic shape of the skyrmion.\\
\begin{figure*}[t]
\includegraphics[width=\textwidth]{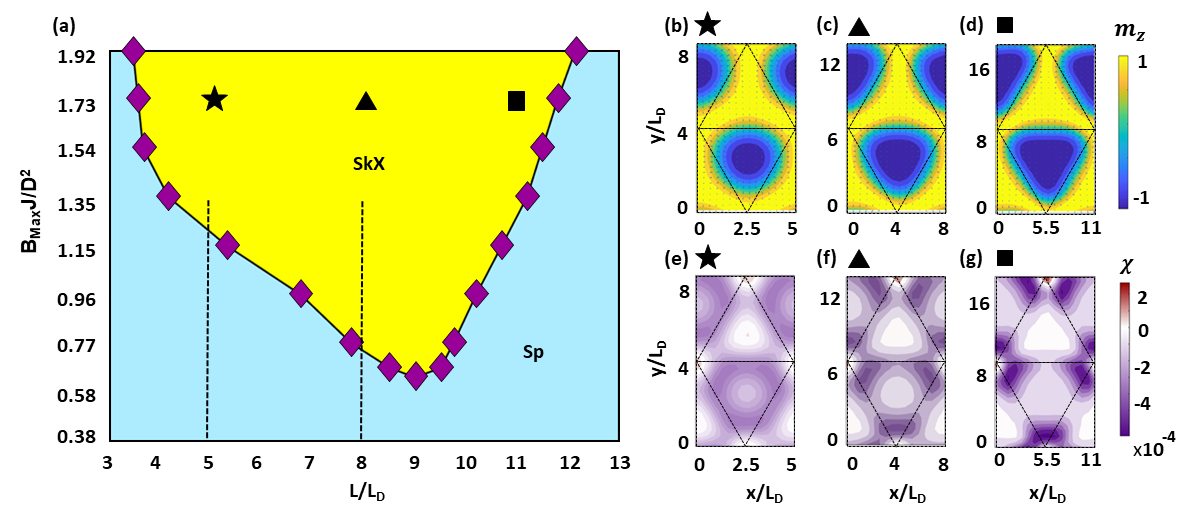}
\caption{$A=0$ phase diagram at different moir\'e periods and exchange field. (a) Moir\'e period $L$ vs  maximum interlayer exchange field $B_{Max}$ phase diagram with spiral (Sp) and skyrmion crystal (SkX) phases. Here $L_D=(J/D)a$, where $a$ is the lattice constant and the purple diamonds correspond to error bars. The dashed lines at $L=5L_D$ and $L=8L_D$ represent the cross section of Fig.~\ref{A_vs_B_{Max}}(a) and Fig.~\ref{A_vs_B(L=5L_D)} respectively. (b,c,d) Magnetization texture and (e,f,g) topological charge density in $1\times 2$ moir\'e super cells for $L/L_D = (5, 8, 11)$ and $B_{Max}J/D^2=1.73$ as marked by $\bigstar, \blacktriangle, \blacksquare$ symbols in (a). The colors represent out of plane component ($m_z$) of magnetization and topological charge density ($\chi$) in (b,c,d) and (e,f,g) respectively. The arrows show the in-plane component of magnetization in (b,c,d) and the dotted lines represent $1\times 2$ moir\'e super cells in (b,c,d,e,f,g).
}
\label{L_vs_B_{Max}}
\end{figure*}

\section{Model}
Before we delve into the analysis of the effective magnetic Hamiltonian, we first describe our setup. As mentioned above, we follow the procedure of Ref. \onlinecite{tong2018skyrmions} to derive the effective interlayer exchange field. We consider a FM monolayer twisted on top of an AFM substrate,  both on a honeycomb lattice with the same lattice constant, $a$. For twisting angle $\theta$, the moir\'e period is given by $L= a/2\sin(\theta/2)$. For small angle $\theta$ and/or lattice mismatch $\delta$, $L\thickapprox a/\sqrt{\theta^2+\delta^2}$ (large period), the local atomic registries on length scale smaller than $L$ but larger than $a$ matches the atomic stacking of different interlayer translation $\textbf{r}$ as shown in Fig.~\ref{Layer+substrate}(c). Hence moir\'e superlattice can be described by interlayer translation vector $\textbf{r(R)}$ that gives atomic registry at position $\textbf{R}$. The interlayer exchange coupling between AFM substrate and FM layer is different at different positions due to different atomic stacking of monolayer and substrate and this leads to spatially dependent exchange field $\textbf{B}\textbf{(R)}$ as shown in Fig.~\ref{Layer+substrate}(b). For example, at position $\textbf{R}_1$ the coupling aligns the spins of 2D layer (green) in positive z-direction when it sits on top of AFM sublattice with spins up (black) and the spins align in opposite direction  at $\textbf{R}_2$ when it is on top of AFM sublattice with spins down (red). The interlayer exchange field at interlayer translation $\textbf{r}$ is give by~\cite{tong2018skyrmions}
\begin{figure*}[t]
\includegraphics[width=\textwidth]{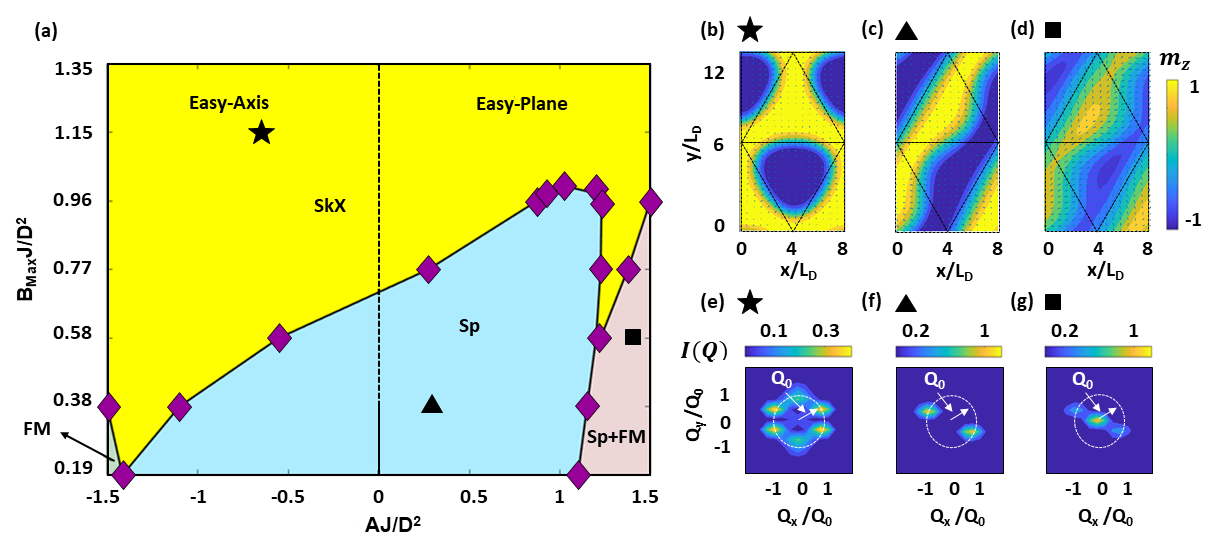}
\caption{$L=8L_D$ phase diagram. (a) Anisotropy $A$ versus maximum interlayer exchange field $B_{Max}$ phase diagram with ferromagnetic (FM), spiral (Sp), skyrmion crystal (SkX) and mixed state (Sp+FM). Here $L_D=(J/D)a$, where $a$ is the lattice constant. (b,c,d) Magnetization texture in $1\times 2$ moir\'e supercells and (e,f,g) spin structure factor for ($\bigstar, \blacktriangle, \blacksquare$) symbols marked in (a). The parameters corresponding to these symbols are the following: $\bigstar$ = $\{ A=-0.6D^2/J$, $B_{Max}=1.15D^2/J \}$, $ \blacktriangle$ = $\{ A=0.3D^2/J$, $B_{Max}=0.38D^2/J \}$ and $\blacksquare$ = $\{ A=1.14D^2/J$, $B_{Max}=0.58D^2/J \}$.
}
\label{A_vs_B_{Max}}
\end{figure*}
\begin{eqnarray}
\textbf{B}^{\tau}(\textbf{r})=\sum_{j,\tau^\prime}J^{\tau^\prime}_{ex}(\textbf{r}-\bm{\xi}_{\tau}+\textbf{R}_j)\textbf{m}_{j,\tau^\prime}
\end{eqnarray}
where $J_{ex}$ is the interlayer coupling coefficient, $\textbf{m}$ is the magnetic moment of top most layer of AFM substrate,  $\left\lbrace\tau,\tau^\prime\right\rbrace=\left\lbrace A,B \right\rbrace$ represents the two in-equivalent sites in unit cell, $\bm{\xi}_A=0$, $\bm{\xi}_B=\left\lbrace 0,a \right\rbrace$ and the summation $j$ is over the Bravais lattice. The total interlayer exchange field per unit cell is given by summing the fields of site $A$ and $B$
\begin{eqnarray}
\textbf{B}(\textbf{r})=\textbf{B}^{A}(\textbf{r})+\textbf{B}^{B}(\textbf{r}).
\end{eqnarray} 
This approximation holds when the interlayer coupling, $J_{ex}$ is small as compared to intralayer coupling, $J$. In general,  $J_{ex}< J$ holds for vdW magnets as the interplane exchange is expected to be much smaller than intraplane exchange.
We used the following coupling form that decays exponentially at long distances 
\begin{eqnarray}
J^{\tau^\prime}_{ex}(\textbf{r}-\bm{\xi}_{\tau}+\textbf{R}_j)=J^{0}_{ex}e^{-\frac{\sqrt{(\textbf{r}-\bm{\xi}_{\tau}+\bm{\xi}_{\tau^\prime}+\textbf{R}_j)^2+d^2}}{r_0}}
\end{eqnarray}
where $d$ is the interlayer separation and $r_{0}$ is the decay length. In our calculations we used $d=a$, $r_0=a$ and this leads to 
\begin{eqnarray}
B_{Max}\thickapprox 1.9244\times 10^{-2}J^{0}_{ex}.
\end{eqnarray} 

Next we describe our model for the monolayer. We consider a magnetic model for 2D honeycomb lattice which is relevant to 2D vdW magnets such as trihallides~\cite{huang2017layer}

\begin{eqnarray} 
H&=& -J\sum_{{r,\mu}} {\bf S}_{r}. ( {\bf S}_{r+\hat{\delta}_{\mu}} )
-D \sum_{{r,\mu}}[
\hat{d}_{\mu}. ({\bf S}_{r} \times {\bf S}_{r+\hat{\delta}_{\mu}} )]
\nonumber\\
&& -A_{c} \sum_{{r,\mu}}[
({\bf S}_{r}.  \hat{d}_{\mu} ) ({\bf S}_{r+\hat{\delta}_{\mu}}.  \hat{d}_{\mu} )
]+A_{s} \sum_{{r}}(S_{r}^z)^2\nonumber\\
&&-\sum_r {\bf B}({\bf R}_r)\cdot {\bf S}_r
\label{Eq:Gnrl_Hmlt_1}
\end{eqnarray}
where $\vec{S}_{r}$ is local moment at site $r$ and $\hat{\delta}_{\mu}$ are the three nearest neighbors on the honeycomb lattice.
$J$ is the ferromagnetic Heisenberg exchange coupling, $D$ is the DM coupling~\cite{Moriya_PR1960}. $A_{c}$ and $A_s $ are the compass and single-ion anisotropies respectively. The DM vector $\hat{d}_{i}=\hat{z}\times\hat{\delta}_{i}$ is set by the symmetry and originates due to the inversion symmetry breaking on the surface. However, the microscopic mechanism for the DM interaction in insulating vdW magnets is different than metallic multilayers since the former is due to the superexchange interaction in the presence of an electric field as originally discussed by Moriya\cite{Moriya_PR1960} whereas the latter can be ascribed to asymmetric interaction paths proposed by Fert and Levy\cite{Levy_PRB1981, Fert_PRL1980}.
$\bf B(R)$ is the interlayer exchange field with the twisted substrate.
To explore the phase diagram of $H$, we consider the free energy functional in the continuum $\mathcal{F}[{\bf m}] = \int d^2r F({\bf m})$ where ${\bf m}({\bf r})$ is the local magnetization.  We set the lattice constant, $a=1$. The free energy density has the following four components
\begin{eqnarray}
F({\bf m}) =F_{iso}+F_{DM}+F_{aniso}+F_{moire}
\end{eqnarray}
where 
\begin{eqnarray}
F_{iso}&=&F_{0}(\textbf{m})+\frac{3}{2}(J/2)\sum_{{\alpha}}(\nabla m^{\alpha})^{2}
\end{eqnarray}
\begin{eqnarray}
F_{DM}&=&-\frac{3}{2}D(m^{z}\partial_{x}m^{x}-m^{x}\partial_{x}m^{z})\nonumber\\
&& +\frac{3}{2}D(m^{y}\partial_{y}m^{z}-m^{z}\partial_{y}m^{y})
\end{eqnarray}
\begin{eqnarray}
F_{aniso}&=&-\frac{3}{2} A_{c} [(m^{x})^{2}+(m^{y})^{2}]+A_{s}(m^{z})^{2}\nonumber\\
&&+\frac{3A_{c}}{4} [ m^{y} (\partial_{x}m^{y}+\partial_{y}m^{x})\nonumber\\
&&-m^{x} (\partial_{x}m^{x}-\partial_{y}m^{y}) ]\\
F_{moire}&=&-{\bf B(R)} \cdot {\bf m}.
\end{eqnarray}
We absorb factor $\frac{3}{2}$ in $J$, $D$ and $A_{c}$ and define the effective anisotropy $A=A_{c}+A_{s}$ which can be positive (easy plane) or negative (easy axis or Ising). The last two terms in $F_{aniso}$, given as $\sim[ m^{y} (\partial_{x}m^{y}+\partial_{y}m^{x}) -m^{x} (\partial_{x}m^{x}-\partial_{y}m^{y}) ]$ have no contribution to the free energy for systems with periodic boundary conditions since they are total derivatives. To obtain the ground state spin configuration $\textbf{m}$, we solve the coupled Landau-Lifshitz-Gilbert (LLG) equations~\cite{1353448}
\begin{eqnarray} 
\frac{d\textbf{m}}{dt}=-\gamma \textbf{m}\times \textbf{B}^{eff}+\alpha \textbf{m} \times \frac{d\textbf{m}}{dt},
\end{eqnarray}
where  $\textbf{B}^{eff}=-\delta H /\delta \textbf{m}$, $\gamma$ is gyromagnetic ratio and $\alpha$ is Gilbert damping coefficient. We start from different initial states\footnote{We consider multiple random configurations as well as ferromagnetic states pointing along (0,0,1), (1,1,5), (1,1,4), (1,1,3), (1,1,2), (1,1,1), (1,1,0), (2,2,1), (3,3,1), (4,4,1), (-1,-1,0), (-1,-1,-1), (-1,-1,-2), (-1,-1,-3), (0,0,-1) directions.} and compare the energies of final states to get the actual ground state. To solve LLG equations  numerically we used mid point method~\cite{d2005numerical} by discretizing the effective magnetic field $\textbf{B}^{eff}=-\delta H /\delta \textbf{m}$ on a $1\times2$ moir\'e supercells. We used $L_D=(J/D)a=10a$ to construct the phase diagrams and the results were also verified at various points for larger values of $L_D$. The magnitude of the magnetization was kept constant at each grid point after each time step enforcing the hard spin constraint, $|{\bf m}|^2=1$ and periodic boundary conditions were imposed at the boundaries. There is unavoidable error due to solving the LLG equations on a discrete lattice. The error is small as long as the variation of the magnetization is small within neighboring sites. We checked that our results are robust as a function of increasing system.

Our method can only capture magnetic orders that are commensurate with the 1x2 moir\'e supercell. However, in the limit of $J_{ex} \rightarrow 0$, the ground state is a spiral with the wave vector $q \simeq D/J$ for weak anisotropy\cite{banerjee2014enhanced} which is in general not commensurate with the moir\'e supercell. Therefore, our method is not suitable to explore the weak $J_{ex}$ limit. Such incommensurate phases and incommensurate to commensurate phase transitions have recently been analyzed in Ref. \citenum{Hejazi_arXiv2020}. The parameters used in this article lie outside the regime of incommensurate phases\cite{Hejazi_arXiv2020}, yet it is not possible to extrapolate our results to $J_{ex} \rightarrow 0$ limit due to this reason.


\section{Results}
We start by exploring the interplay between the moir\'e periodicity $L$ and $B_{Max}$ as shown in Fig.~\ref{L_vs_B_{Max}} for $A=0$. $B_{Max}$ is the maximum value of interlayer exchange magnetic field. As shown in Fig 1(b), the moir\'e supercell splits into two triangles of opposite alternating effective magnetic field which is maximum at the centers and vanishingly small at the corners of the triangles. Therefore, in the limit of large $B_{Max}$, the magnetization aligns with the effective magnetic field at the center of the triangles, creating two ferromagnetic domains separated with a domain wall whose chirality is set by D. The only degree of freedom left that determines the ground state is the magnetization at the corners of the triangles which creates an effective triangular lattice. In particular, if the magnetization at the corners are ferromagnetic, the state is a skyrmion whereas a stripe order give rise to a spiral phase. We find that the ground state is a spiral for low exchange field. As we increase the field, the SkX phase starts at the moir\'e period $L/L_{D}\approx9$ which corresponds to the optimum angle between the layer and the  substrate. As we further increase the field, we get SkX for a range of values around optimum period. This range increases with increasing the exchange field. Unlike skyrmions in chiral magnets where the size of the skyrmion is set by $L_D$, here we find that the size of skyrmions is determined by the moir\'e period. On the other hand, $L_D$ determines the boundary length between the interior and the exterior of the skyrmions. For small moir\'e period, skyrmions are small and their shape is nearly circular as shown in Fig.~\ref{L_vs_B_{Max}}(b). Fig~.\ref{L_vs_B_{Max}}(c) shows that as the period increases, the size of skyrmion also increases and it takes the triangular shape of the exchange field. The corners of skyrmions get sharper with increasing $L$. Unlike the skyrmions in chiral magnets, we find that a large fraction of the topological charge is concentrated at the edges of the skyrmion. This fraction increases with increasing $L$. There is also a small fraction of opposite charge between the skyrmions which decreases with increasing $L$. This charge arises due to the anti-vortices between the skyrmions\cite{lin2015skyrmion}. For large $L$, the topological charge further splits into three parts due to the triangular shape of the skyrmion as shown in Fig.~\ref{L_vs_B_{Max}}(g).\\

\begin{figure}
\includegraphics[width=\columnwidth]{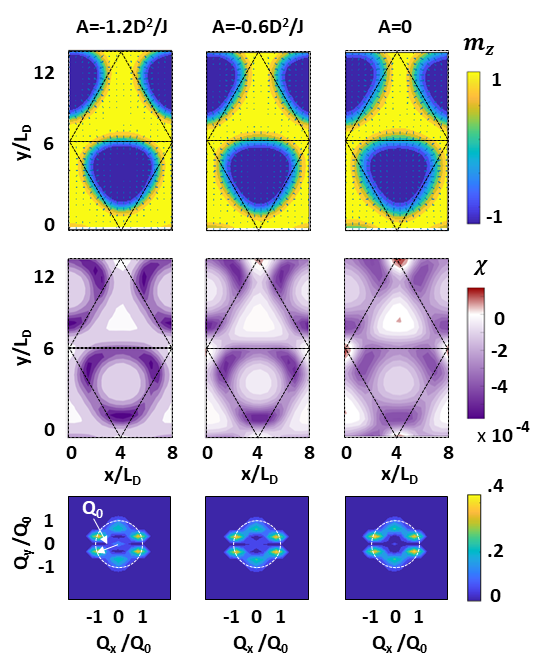}
\caption{Evolution of magnetization texture, topological charge density $\chi$ and spin structure factor $I(Q)$ for $AJ/D^2=\{-1.2, -0.6, 0\}$ at $L=8L_D$.}
\label{SkX_vs_A}
\end{figure}
Next, we explore the effects of anisotropy, $A$ at around optimal angle $L=8L_{D}$ as shown in Fig.~\ref{A_vs_B_{Max}}. At low exchange field, we obtain spiral phase for a wide range of $A$, a small ferromagnetic phase at lowest negative values of $A$ as well as a mixed (FM+Sp) phase at largest positive values of $A$. As the field increases, initially we get SkX near the lowest negative values of $A$ that corresponds to easy axis anisotropy. By further increasing the field, the range of SkX gradually increases and eventually, occupies the whole phase diagram. Fig.~\ref{A_vs_B_{Max}}(b,c,d) show the local magnetization and Fig.~\ref{A_vs_B_{Max}}(e,f,g) show the spin structure factor $I(Q)\propto|\langle\textbf{m}_{Q}\rangle|^2$ for the three phases. $\textbf{m}_Q$ is the Fourier transform of the 
magnetization. Unlike an isotropic SkX which has six peaks on the circle in the spin structure factor, we find four peaks lie on circle and 
two peaks lie inside the circle. This is due to the anisotropic triangular shape of the SkX. The spiral has two peaks at $Q = \pm Q_0$ and the mixed state (FM+Sp) has three peaks including the $Q=0$ from the FM and $Q=\pm Q_0$ from the spiral phase. The intensity of $Q=0$ 
peak increases with increasing $A$ and decreases with increasing exchange field $B_{Max}$. On the other hand, the intensities corresponding to the spiral wave vectors have the opposite behavior of $Q=0$ with $A$ and $B_{Max}$. \\
The properties of SkX also depends on the anisotropy $A$. Fig.~\ref{SkX_vs_A} shows the magnetization, spin structure factor and topological charge density as a function of $A$ for moir\'e period $L=8L_D$. For $A=-1.2D^2/J$ the skyrmion has a sharp boundary wall 
where magnetization changes abruptly and then it changes slowly inside the skyrmion. The sharpness of boundary wall decreases with 
increasing $A$ and the change in magnetization inside the skyrmion increases with increasing $A$. For $A=-1.2D^2/J$, a large fraction of topological charge is concentrated at the boundary wall of skyrmion and a small fraction lies inside the skyrmion. There is also a small fraction of opposite charge between the skyrmions. The concentration of charge at boundary decreases with increasing $A$ and the central charge increases with increasing $A$. The fraction of opposite charge between the skyrmions also increases with increasing $A$\cite{lin2015skyrmion}.

\begin{figure}[h]
\includegraphics[width=\columnwidth]{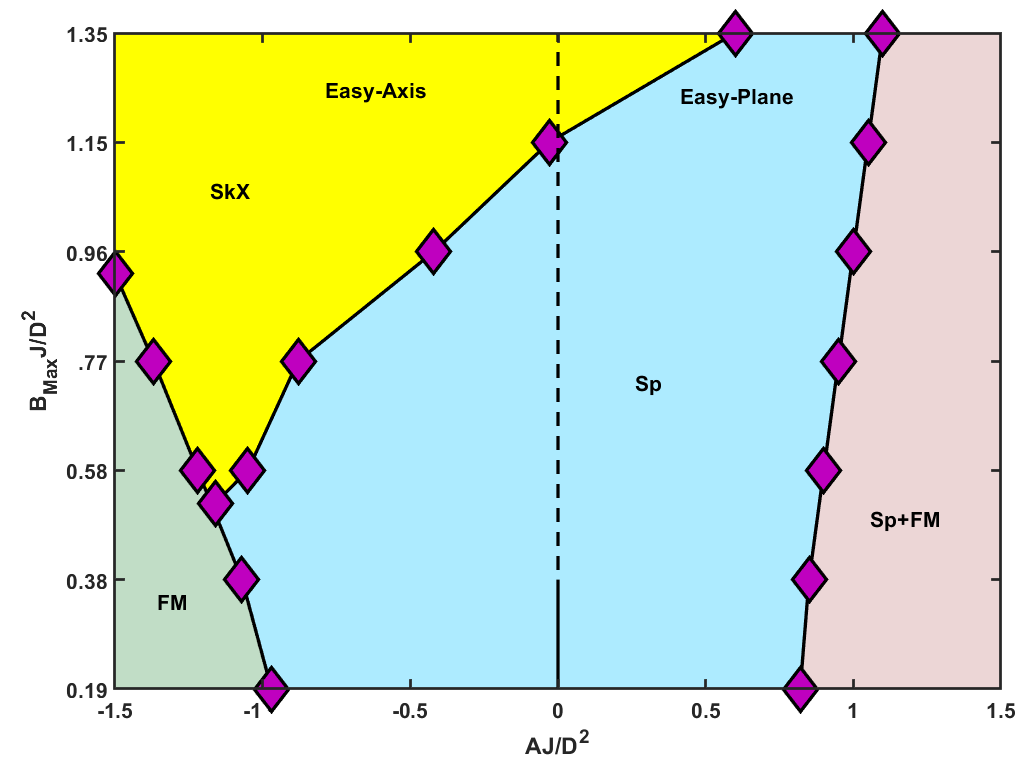}
\caption{Anisotropy $A$ vs  maximum interlayer exchange field $B_{Max}$ phase diagram with ferromagnetic (FM), spiral (Sp), skyrmion crystal (SkX) and mixed state (Sp+FM) phases at $L=5L_D$. Here $L_D=(J/D)a$, where $a$ is the lattice constant.}
\label{A_vs_B(L=5L_D)}
\end{figure}

We also studied the effects of anisotropy for a non-optimal angle at $L = 5L_D$. As shown in Fig.~\ref{A_vs_B(L=5L_D)}, SkX is highly suppressed in this case but still persists for large exchange field and easy axis anisotropy. Suppression of SkX is due to the fact that the moir\'e supercell is too small with respect to the optimal size of the skyrmions.

Our results apply to a wide range of vdW magnets since the primary constraint in our model is the requirement that $J_{ex}<J$ which is in general satisfied for vdW magnets. Our phase diagrams span a wide range of parameters and for a fixed twisting angle, our results scale with $AJ/D^2$.

\section{Conclusion}
We have shown that skyrmion crystal can be stabilized in moir\'e superlattices in the absence of external magnetic field. We found a large SkX phase for easy axis anisotropy which can be essential to stabilize skyrmions in vdW magnets such as CrI$_3$\cite{huang2017layer}. In particular, for optimal moir\'e periodicity SkX occupies the majority of the phase diagram. We find that the properties of the skyrmion can be tuned with the moir\'e periodicity and anisotropy. Unlike skyrmions in chiral magnets, the topological charge density depends on the size of the skyrmions sand it is concentrated at the edges for skyrmions with large L.

\section{Acknowledgements}
We thank Sumilan Banerjee for useful discussions. MA is supported by Fulbright scholarship and ASU startup grant and OE is supported by NSF-DMR-1904716. We acknowledge the ASU Research Computing Center for HPC resources.



\begin{thebibliography}{49}
\expandafter\ifx\csname natexlab\endcsname\relax\def\natexlab#1{#1}\fi
\expandafter\ifx\csname bibnamefont\endcsname\relax
  \def\bibnamefont#1{#1}\fi
\expandafter\ifx\csname bibfnamefont\endcsname\relax
  \def\bibfnamefont#1{#1}\fi
\expandafter\ifx\csname citenamefont\endcsname\relax
  \def\citenamefont#1{#1}\fi
\expandafter\ifx\csname url\endcsname\relax
  \def\url#1{\texttt{#1}}\fi
\expandafter\ifx\csname urlprefix\endcsname\relax\def\urlprefix{URL }\fi
\providecommand{\bibinfo}[2]{#2}
\providecommand{\eprint}[2][]{\url{#2}}

\bibitem[{\citenamefont{Gibertini et~al.}(2019)\citenamefont{Gibertini,
  Koperski, Morpurgo, and Novoselov}}]{gibertini2019magnetic}
\bibinfo{author}{\bibfnamefont{M.}~\bibnamefont{Gibertini}},
  \bibinfo{author}{\bibfnamefont{M.}~\bibnamefont{Koperski}},
  \bibinfo{author}{\bibfnamefont{A.}~\bibnamefont{Morpurgo}}, \bibnamefont{and}
  \bibinfo{author}{\bibfnamefont{K.}~\bibnamefont{Novoselov}},
  \bibinfo{journal}{Nature Nanotechnology} \textbf{\bibinfo{volume}{14}},
  \bibinfo{pages}{408} (\bibinfo{year}{2019}).

\bibitem[{\citenamefont{Gong and Zhang}(2019)}]{gong2019two}
\bibinfo{author}{\bibfnamefont{C.}~\bibnamefont{Gong}} \bibnamefont{and}
  \bibinfo{author}{\bibfnamefont{X.}~\bibnamefont{Zhang}},
  \bibinfo{journal}{Science} \textbf{\bibinfo{volume}{363}},
  \bibinfo{pages}{eaav4450} (\bibinfo{year}{2019}).

\bibitem[{\citenamefont{Burch et~al.}(2018)\citenamefont{Burch, Mandrus, and
  Park}}]{burch2018magnetism}
\bibinfo{author}{\bibfnamefont{K.~S.} \bibnamefont{Burch}},
  \bibinfo{author}{\bibfnamefont{D.}~\bibnamefont{Mandrus}}, \bibnamefont{and}
  \bibinfo{author}{\bibfnamefont{J.-G.} \bibnamefont{Park}},
  \bibinfo{journal}{Nature} \textbf{\bibinfo{volume}{563}}, \bibinfo{pages}{47}
  (\bibinfo{year}{2018}).

\bibitem[{\citenamefont{Huang et~al.}(2017)\citenamefont{Huang, Clark,
  Navarro-Moratalla, Klein, Cheng, Seyler, Zhong, Schmidgall, McGuire, Cobden
  et~al.}}]{huang2017layer}
\bibinfo{author}{\bibfnamefont{B.}~\bibnamefont{Huang}},
  \bibinfo{author}{\bibfnamefont{G.}~\bibnamefont{Clark}},
  \bibinfo{author}{\bibfnamefont{E.}~\bibnamefont{Navarro-Moratalla}},
  \bibinfo{author}{\bibfnamefont{D.~R.} \bibnamefont{Klein}},
  \bibinfo{author}{\bibfnamefont{R.}~\bibnamefont{Cheng}},
  \bibinfo{author}{\bibfnamefont{K.~L.} \bibnamefont{Seyler}},
  \bibinfo{author}{\bibfnamefont{D.}~\bibnamefont{Zhong}},
  \bibinfo{author}{\bibfnamefont{E.}~\bibnamefont{Schmidgall}},
  \bibinfo{author}{\bibfnamefont{M.~A.} \bibnamefont{McGuire}},
  \bibinfo{author}{\bibfnamefont{D.~H.} \bibnamefont{Cobden}},
  \bibnamefont{et~al.}, \bibinfo{journal}{Nature}
  \textbf{\bibinfo{volume}{546}}, \bibinfo{pages}{270} (\bibinfo{year}{2017}).
  
\bibitem[{\citenamefont{Moriya}(1960{\natexlab{b}})}]{Moriya_PR1960}
\bibinfo{author}{\bibfnamefont{T.}~\bibnamefont{Moriya}},
  \bibinfo{journal}{Phys. Rev.} \textbf{\bibinfo{volume}{120}},
  \bibinfo{pages}{91} (\bibinfo{year}{1960}{\natexlab{b}}).

\bibitem[{\citenamefont{Gong et~al.}(2017)\citenamefont{Gong, Li, Li, Ji,
  Stern, Xia, Cao, Bao, Wang, Wang et~al.}}]{gong2017discovery}
\bibinfo{author}{\bibfnamefont{C.}~\bibnamefont{Gong}},
  \bibinfo{author}{\bibfnamefont{L.}~\bibnamefont{Li}},
  \bibinfo{author}{\bibfnamefont{Z.}~\bibnamefont{Li}},
  \bibinfo{author}{\bibfnamefont{H.}~\bibnamefont{Ji}},
  \bibinfo{author}{\bibfnamefont{A.}~\bibnamefont{Stern}},
  \bibinfo{author}{\bibfnamefont{Y.}~\bibnamefont{Xia}},
  \bibinfo{author}{\bibfnamefont{T.}~\bibnamefont{Cao}},
  \bibinfo{author}{\bibfnamefont{W.}~\bibnamefont{Bao}},
  \bibinfo{author}{\bibfnamefont{C.}~\bibnamefont{Wang}},
  \bibinfo{author}{\bibfnamefont{Y.}~\bibnamefont{Wang}}, \bibnamefont{et~al.},
  \bibinfo{journal}{Nature} \textbf{\bibinfo{volume}{546}},
  \bibinfo{pages}{265} (\bibinfo{year}{2017}).

\bibitem[{\citenamefont{Deng et~al.}(2018)\citenamefont{Deng, Yu, Song, Zhang,
  Wang, Sun, Yi, Wu, Wu, Zhu et~al.}}]{deng2018gate}
\bibinfo{author}{\bibfnamefont{Y.}~\bibnamefont{Deng}},
  \bibinfo{author}{\bibfnamefont{Y.}~\bibnamefont{Yu}},
  \bibinfo{author}{\bibfnamefont{Y.}~\bibnamefont{Song}},
  \bibinfo{author}{\bibfnamefont{J.}~\bibnamefont{Zhang}},
  \bibinfo{author}{\bibfnamefont{N.~Z.} \bibnamefont{Wang}},
  \bibinfo{author}{\bibfnamefont{Z.}~\bibnamefont{Sun}},
  \bibinfo{author}{\bibfnamefont{Y.}~\bibnamefont{Yi}},
  \bibinfo{author}{\bibfnamefont{Y.~Z.} \bibnamefont{Wu}},
  \bibinfo{author}{\bibfnamefont{S.}~\bibnamefont{Wu}},
  \bibinfo{author}{\bibfnamefont{J.}~\bibnamefont{Zhu}}, \bibnamefont{et~al.},
  \bibinfo{journal}{Nature} \textbf{\bibinfo{volume}{563}}, \bibinfo{pages}{94}
  (\bibinfo{year}{2018}).

\bibitem[{\citenamefont{Bonilla et~al.}(2018)\citenamefont{Bonilla, Kolekar,
  Ma, Diaz, Kalappattil, Das, Eggers, Gutierrez, Phan, and
  Batzill}}]{bonilla2018strong}
\bibinfo{author}{\bibfnamefont{M.}~\bibnamefont{Bonilla}},
  \bibinfo{author}{\bibfnamefont{S.}~\bibnamefont{Kolekar}},
  \bibinfo{author}{\bibfnamefont{Y.}~\bibnamefont{Ma}},
  \bibinfo{author}{\bibfnamefont{H.~C.} \bibnamefont{Diaz}},
  \bibinfo{author}{\bibfnamefont{V.}~\bibnamefont{Kalappattil}},
  \bibinfo{author}{\bibfnamefont{R.}~\bibnamefont{Das}},
  \bibinfo{author}{\bibfnamefont{T.}~\bibnamefont{Eggers}},
  \bibinfo{author}{\bibfnamefont{H.~R.} \bibnamefont{Gutierrez}},
  \bibinfo{author}{\bibfnamefont{M.-H.} \bibnamefont{Phan}}, \bibnamefont{and}
  \bibinfo{author}{\bibfnamefont{M.}~\bibnamefont{Batzill}},
  \bibinfo{journal}{Nature Nanotechnology} \textbf{\bibinfo{volume}{13}},
  \bibinfo{pages}{289} (\bibinfo{year}{2018}).

\bibitem[{\citenamefont{O'Hara et~al.}(2018)\citenamefont{O'Hara, Zhu,
  Trout, Ahmed, Luo, Lee, Brenner, Rajan, Gupta, McComb et~al.}}]{o2018room}
\bibinfo{author}{\bibfnamefont{D.~J.} \bibnamefont{O'Hara}},
  \bibinfo{author}{\bibfnamefont{T.}~\bibnamefont{Zhu}},
  \bibinfo{author}{\bibfnamefont{A.~H.} \bibnamefont{Trout}},
  \bibinfo{author}{\bibfnamefont{A.~S.} \bibnamefont{Ahmed}},
  \bibinfo{author}{\bibfnamefont{Y.~K.} \bibnamefont{Luo}},
  \bibinfo{author}{\bibfnamefont{C.~H.} \bibnamefont{Lee}},
  \bibinfo{author}{\bibfnamefont{M.~R.} \bibnamefont{Brenner}},
  \bibinfo{author}{\bibfnamefont{S.}~\bibnamefont{Rajan}},
  \bibinfo{author}{\bibfnamefont{J.~A.} \bibnamefont{Gupta}},
  \bibinfo{author}{\bibfnamefont{D.~W.} \bibnamefont{McComb}},
  \bibnamefont{et~al.}, \bibinfo{journal}{Nano Letters}
  \textbf{\bibinfo{volume}{18}}, \bibinfo{pages}{3125} (\bibinfo{year}{2018}).

\bibitem[{\citenamefont{Soriano et~al.}(2019)\citenamefont{Soriano, Cardoso,
  and Fern{\'a}ndez-Rossier}}]{soriano2019interplay}
\bibinfo{author}{\bibfnamefont{D.}~\bibnamefont{Soriano}},
  \bibinfo{author}{\bibfnamefont{C.}~\bibnamefont{Cardoso}}, \bibnamefont{and}
  \bibinfo{author}{\bibfnamefont{J.}~\bibnamefont{Fern{\'a}ndez-Rossier}},
  \bibinfo{journal}{Solid State Communications} \textbf{\bibinfo{volume}{299}},
  \bibinfo{pages}{113662} (\bibinfo{year}{2019}).

\bibitem[{\citenamefont{Cristoloveanu et~al.}(2019)\citenamefont{Cristoloveanu,
  Lee, Park, and Parihar}}]{cristoloveanu2019concept}
\bibinfo{author}{\bibfnamefont{S.}~\bibnamefont{Cristoloveanu}},
  \bibinfo{author}{\bibfnamefont{K.~H.} \bibnamefont{Lee}},
  \bibinfo{author}{\bibfnamefont{H.}~\bibnamefont{Park}}, \bibnamefont{and}
  \bibinfo{author}{\bibfnamefont{M.~S.} \bibnamefont{Parihar}},
  \bibinfo{journal}{Solid-State Electronics} \textbf{\bibinfo{volume}{155}},
  \bibinfo{pages}{32} (\bibinfo{year}{2019}).

\bibitem[{\citenamefont{Ye et~al.}(2012)\citenamefont{Ye, Zhang, Akashi,
  Bahramy, Arita, and Iwasa}}]{ye2012superconducting}
\bibinfo{author}{\bibfnamefont{J.}~\bibnamefont{Ye}},
  \bibinfo{author}{\bibfnamefont{Y.~J.} \bibnamefont{Zhang}},
  \bibinfo{author}{\bibfnamefont{R.}~\bibnamefont{Akashi}},
  \bibinfo{author}{\bibfnamefont{M.~S.} \bibnamefont{Bahramy}},
  \bibinfo{author}{\bibfnamefont{R.}~\bibnamefont{Arita}}, \bibnamefont{and}
  \bibinfo{author}{\bibfnamefont{Y.}~\bibnamefont{Iwasa}},
  \bibinfo{journal}{Science} \textbf{\bibinfo{volume}{338}},
  \bibinfo{pages}{1193} (\bibinfo{year}{2012}).

\bibitem[{\citenamefont{Cao et~al.}(2018{\natexlab{a}})\citenamefont{Cao,
  Fatemi, Fang, Watanabe, Taniguchi, Kaxiras, and
  Jarillo-Herrero}}]{cao2018unconventional}
\bibinfo{author}{\bibfnamefont{Y.}~\bibnamefont{Cao}},
  \bibinfo{author}{\bibfnamefont{V.}~\bibnamefont{Fatemi}},
  \bibinfo{author}{\bibfnamefont{S.}~\bibnamefont{Fang}},
  \bibinfo{author}{\bibfnamefont{K.}~\bibnamefont{Watanabe}},
  \bibinfo{author}{\bibfnamefont{T.}~\bibnamefont{Taniguchi}},
  \bibinfo{author}{\bibfnamefont{E.}~\bibnamefont{Kaxiras}}, \bibnamefont{and}
  \bibinfo{author}{\bibfnamefont{P.}~\bibnamefont{Jarillo-Herrero}},
  \bibinfo{journal}{Nature} \textbf{\bibinfo{volume}{556}}, \bibinfo{pages}{43}
  (\bibinfo{year}{2018}{\natexlab{a}}).

\bibitem[{\citenamefont{Huang et~al.}(2018)\citenamefont{Huang, Clark, Klein,
  MacNeill, Navarro-Moratalla, Seyler, Wilson, McGuire, Cobden, Xiao
  et~al.}}]{huang2018electrical}
\bibinfo{author}{\bibfnamefont{B.}~\bibnamefont{Huang}},
  \bibinfo{author}{\bibfnamefont{G.}~\bibnamefont{Clark}},
  \bibinfo{author}{\bibfnamefont{D.~R.} \bibnamefont{Klein}},
  \bibinfo{author}{\bibfnamefont{D.}~\bibnamefont{MacNeill}},
  \bibinfo{author}{\bibfnamefont{E.}~\bibnamefont{Navarro-Moratalla}},
  \bibinfo{author}{\bibfnamefont{K.~L.} \bibnamefont{Seyler}},
  \bibinfo{author}{\bibfnamefont{N.}~\bibnamefont{Wilson}},
  \bibinfo{author}{\bibfnamefont{M.~A.} \bibnamefont{McGuire}},
  \bibinfo{author}{\bibfnamefont{D.~H.} \bibnamefont{Cobden}},
  \bibinfo{author}{\bibfnamefont{D.}~\bibnamefont{Xiao}}, \bibnamefont{et~al.},
  \bibinfo{journal}{Nature Nanotechnology} \textbf{\bibinfo{volume}{13}},
  \bibinfo{pages}{544} (\bibinfo{year}{2018}).

\bibitem[{\citenamefont{Song et~al.}(2019)\citenamefont{Song, Fei, Yankowitz,
  Lin, Jiang, Hwangbo, Zhang, Sun, Taniguchi, Watanabe
  et~al.}}]{song2019switching}
\bibinfo{author}{\bibfnamefont{T.}~\bibnamefont{Song}},
  \bibinfo{author}{\bibfnamefont{Z.}~\bibnamefont{Fei}},
  \bibinfo{author}{\bibfnamefont{M.}~\bibnamefont{Yankowitz}},
  \bibinfo{author}{\bibfnamefont{Z.}~\bibnamefont{Lin}},
  \bibinfo{author}{\bibfnamefont{Q.}~\bibnamefont{Jiang}},
  \bibinfo{author}{\bibfnamefont{K.}~\bibnamefont{Hwangbo}},
  \bibinfo{author}{\bibfnamefont{Q.}~\bibnamefont{Zhang}},
  \bibinfo{author}{\bibfnamefont{B.}~\bibnamefont{Sun}},
  \bibinfo{author}{\bibfnamefont{T.}~\bibnamefont{Taniguchi}},
  \bibinfo{author}{\bibfnamefont{K.}~\bibnamefont{Watanabe}},
  \bibnamefont{et~al.}, \bibinfo{journal}{Nature Materials}
  \textbf{\bibinfo{volume}{18}}, \bibinfo{pages}{1} (\bibinfo{year}{2019}).

\bibitem[{\citenamefont{Ponomarenko et~al.}(2013)\citenamefont{Ponomarenko,
  Gorbachev, Yu, Elias, Jalil, Patel, Mishchenko, Mayorov, Woods, Wallbank
  et~al.}}]{ponomarenko2013cloning}
\bibinfo{author}{\bibfnamefont{L.}~\bibnamefont{Ponomarenko}},
  \bibinfo{author}{\bibfnamefont{R.}~\bibnamefont{Gorbachev}},
  \bibinfo{author}{\bibfnamefont{G.}~\bibnamefont{Yu}},
  \bibinfo{author}{\bibfnamefont{D.}~\bibnamefont{Elias}},
  \bibinfo{author}{\bibfnamefont{R.}~\bibnamefont{Jalil}},
  \bibinfo{author}{\bibfnamefont{A.}~\bibnamefont{Patel}},
  \bibinfo{author}{\bibfnamefont{A.}~\bibnamefont{Mishchenko}},
  \bibinfo{author}{\bibfnamefont{A.}~\bibnamefont{Mayorov}},
  \bibinfo{author}{\bibfnamefont{C.}~\bibnamefont{Woods}},
  \bibinfo{author}{\bibfnamefont{J.}~\bibnamefont{Wallbank}},
  \bibnamefont{et~al.}, \bibinfo{journal}{Nature}
  \textbf{\bibinfo{volume}{497}}, \bibinfo{pages}{594} (\bibinfo{year}{2013}).

\bibitem[{\citenamefont{Dean et~al.}(2013)\citenamefont{Dean, Wang, Maher,
  Forsythe, Ghahari, Gao, Katoch, Ishigami, Moon, Koshino
  et~al.}}]{dean2013hofstadter}
\bibinfo{author}{\bibfnamefont{C.~R.} \bibnamefont{Dean}},
  \bibinfo{author}{\bibfnamefont{L.}~\bibnamefont{Wang}},
  \bibinfo{author}{\bibfnamefont{P.}~\bibnamefont{Maher}},
  \bibinfo{author}{\bibfnamefont{C.}~\bibnamefont{Forsythe}},
  \bibinfo{author}{\bibfnamefont{F.}~\bibnamefont{Ghahari}},
  \bibinfo{author}{\bibfnamefont{Y.}~\bibnamefont{Gao}},
  \bibinfo{author}{\bibfnamefont{J.}~\bibnamefont{Katoch}},
  \bibinfo{author}{\bibfnamefont{M.}~\bibnamefont{Ishigami}},
  \bibinfo{author}{\bibfnamefont{P.}~\bibnamefont{Moon}},
  \bibinfo{author}{\bibfnamefont{M.}~\bibnamefont{Koshino}},
  \bibnamefont{et~al.}, \bibinfo{journal}{Nature}
  \textbf{\bibinfo{volume}{497}}, \bibinfo{pages}{598} (\bibinfo{year}{2013}).

\bibitem[{\citenamefont{Gorbachev et~al.}(2014)\citenamefont{Gorbachev, Song,
  Yu, Kretinin, Withers, Cao, Mishchenko, Grigorieva, Novoselov, Levitov
  et~al.}}]{gorbachev2014detecting}
\bibinfo{author}{\bibfnamefont{R.}~\bibnamefont{Gorbachev}},
  \bibinfo{author}{\bibfnamefont{J.}~\bibnamefont{Song}},
  \bibinfo{author}{\bibfnamefont{G.}~\bibnamefont{Yu}},
  \bibinfo{author}{\bibfnamefont{A.}~\bibnamefont{Kretinin}},
  \bibinfo{author}{\bibfnamefont{F.}~\bibnamefont{Withers}},
  \bibinfo{author}{\bibfnamefont{Y.}~\bibnamefont{Cao}},
  \bibinfo{author}{\bibfnamefont{A.}~\bibnamefont{Mishchenko}},
  \bibinfo{author}{\bibfnamefont{I.}~\bibnamefont{Grigorieva}},
  \bibinfo{author}{\bibfnamefont{K.}~\bibnamefont{Novoselov}},
  \bibinfo{author}{\bibfnamefont{L.}~\bibnamefont{Levitov}},
  \bibnamefont{et~al.}, \bibinfo{journal}{Science}
  \textbf{\bibinfo{volume}{346}}, \bibinfo{pages}{448} (\bibinfo{year}{2014}).

\bibitem[{\citenamefont{Cao et~al.}(2018{\natexlab{b}})\citenamefont{Cao,
  Fatemi, Demir, Fang, Tomarken, Luo, Sanchez-Yamagishi, Watanabe, Taniguchi,
  Kaxiras et~al.}}]{cao2018correlated}
\bibinfo{author}{\bibfnamefont{Y.}~\bibnamefont{Cao}},
  \bibinfo{author}{\bibfnamefont{V.}~\bibnamefont{Fatemi}},
  \bibinfo{author}{\bibfnamefont{A.}~\bibnamefont{Demir}},
  \bibinfo{author}{\bibfnamefont{S.}~\bibnamefont{Fang}},
  \bibinfo{author}{\bibfnamefont{S.~L.} \bibnamefont{Tomarken}},
  \bibinfo{author}{\bibfnamefont{J.~Y.} \bibnamefont{Luo}},
  \bibinfo{author}{\bibfnamefont{J.~D.} \bibnamefont{Sanchez-Yamagishi}},
  \bibinfo{author}{\bibfnamefont{K.}~\bibnamefont{Watanabe}},
  \bibinfo{author}{\bibfnamefont{T.}~\bibnamefont{Taniguchi}},
  \bibinfo{author}{\bibfnamefont{E.}~\bibnamefont{Kaxiras}},
  \bibnamefont{et~al.}, \bibinfo{journal}{Nature}
  \textbf{\bibinfo{volume}{556}}, \bibinfo{pages}{80}
  (\bibinfo{year}{2018}{\natexlab{b}}).

\bibitem[{\citenamefont{Zhang et~al.}(2017)\citenamefont{Zhang, Chuu, Ren, Li,
  Li, Jin, Chou, and Shih}}]{zhang2017interlayer}
\bibinfo{author}{\bibfnamefont{C.}~\bibnamefont{Zhang}},
  \bibinfo{author}{\bibfnamefont{C.-P.} \bibnamefont{Chuu}},
  \bibinfo{author}{\bibfnamefont{X.}~\bibnamefont{Ren}},
  \bibinfo{author}{\bibfnamefont{M.-Y.} \bibnamefont{Li}},
  \bibinfo{author}{\bibfnamefont{L.-J.} \bibnamefont{Li}},
  \bibinfo{author}{\bibfnamefont{C.}~\bibnamefont{Jin}},
  \bibinfo{author}{\bibfnamefont{M.-Y.} \bibnamefont{Chou}}, \bibnamefont{and}
  \bibinfo{author}{\bibfnamefont{C.-K.} \bibnamefont{Shih}},
  \bibinfo{journal}{Science Advances} \textbf{\bibinfo{volume}{3}},
  \bibinfo{pages}{e1601459} (\bibinfo{year}{2017}).

\bibitem[{\citenamefont{Moon and Koshino}(2014)}]{moon2014electronic}
\bibinfo{author}{\bibfnamefont{P.}~\bibnamefont{Moon}} \bibnamefont{and}
  \bibinfo{author}{\bibfnamefont{M.}~\bibnamefont{Koshino}},
  \bibinfo{journal}{Physical Review B} \textbf{\bibinfo{volume}{90}},
  \bibinfo{pages}{155406} (\bibinfo{year}{2014}).

\bibitem[{\citenamefont{Chen et~al.}(2019{\natexlab{a}})\citenamefont{Chen,
  Jiang, Wu, Lyu, Li, Chittari, Watanabe, Taniguchi, Shi, Jung
  et~al.}}]{chen2019evidence}
\bibinfo{author}{\bibfnamefont{G.}~\bibnamefont{Chen}},
  \bibinfo{author}{\bibfnamefont{L.}~\bibnamefont{Jiang}},
  \bibinfo{author}{\bibfnamefont{S.}~\bibnamefont{Wu}},
  \bibinfo{author}{\bibfnamefont{B.}~\bibnamefont{Lyu}},
  \bibinfo{author}{\bibfnamefont{H.}~\bibnamefont{Li}},
  \bibinfo{author}{\bibfnamefont{B.~L.} \bibnamefont{Chittari}},
  \bibinfo{author}{\bibfnamefont{K.}~\bibnamefont{Watanabe}},
  \bibinfo{author}{\bibfnamefont{T.}~\bibnamefont{Taniguchi}},
  \bibinfo{author}{\bibfnamefont{Z.}~\bibnamefont{Shi}},
  \bibinfo{author}{\bibfnamefont{J.}~\bibnamefont{Jung}}, \bibnamefont{et~al.},
  \bibinfo{journal}{Nature Physics} \textbf{\bibinfo{volume}{15}},
  \bibinfo{pages}{237} (\bibinfo{year}{2019}{\natexlab{a}}).

\bibitem[{\citenamefont{Chen et~al.}(2019{\natexlab{b}})\citenamefont{Chen,
  Sharpe, Gallagher, Rosen, Fox, Jiang, Lyu, Li, Watanabe, Taniguchi
  et~al.}}]{chen2019signatures}
\bibinfo{author}{\bibfnamefont{G.}~\bibnamefont{Chen}},
  \bibinfo{author}{\bibfnamefont{A.~L.} \bibnamefont{Sharpe}},
  \bibinfo{author}{\bibfnamefont{P.}~\bibnamefont{Gallagher}},
  \bibinfo{author}{\bibfnamefont{I.~T.} \bibnamefont{Rosen}},
  \bibinfo{author}{\bibfnamefont{E.~J.} \bibnamefont{Fox}},
  \bibinfo{author}{\bibfnamefont{L.}~\bibnamefont{Jiang}},
  \bibinfo{author}{\bibfnamefont{B.}~\bibnamefont{Lyu}},
  \bibinfo{author}{\bibfnamefont{H.}~\bibnamefont{Li}},
  \bibinfo{author}{\bibfnamefont{K.}~\bibnamefont{Watanabe}},
  \bibinfo{author}{\bibfnamefont{T.}~\bibnamefont{Taniguchi}},
  \bibnamefont{et~al.}, \bibinfo{journal}{Nature}
  \textbf{\bibinfo{volume}{572}}, \bibinfo{pages}{215}
  (\bibinfo{year}{2019}{\natexlab{b}}).

\bibitem[{\citenamefont{Wallbank et~al.}(2015)\citenamefont{Wallbank,
  Mucha-Kruczy{\'n}ski, Chen, and Fal'ko}}]{wallbank2015moire}
\bibinfo{author}{\bibfnamefont{J.~R.} \bibnamefont{Wallbank}},
  \bibinfo{author}{\bibfnamefont{M.}~\bibnamefont{Mucha-Kruczy{\'n}ski}},
  \bibinfo{author}{\bibfnamefont{X.}~\bibnamefont{Chen}}, \bibnamefont{and}
  \bibinfo{author}{\bibfnamefont{V.~I.} \bibnamefont{Fal'ko}},
  \bibinfo{journal}{Annalen der Physik} \textbf{\bibinfo{volume}{527}},
  \bibinfo{pages}{359} (\bibinfo{year}{2015}).

\bibitem[{\citenamefont{Ni et~al.}(2015)\citenamefont{Ni, Wang, Wu, Fei,
  Goldflam, Keilmann, {\"O}zyilmaz, Neto, Xie, Fogler et~al.}}]{ni2015plasmons}
\bibinfo{author}{\bibfnamefont{G.}~\bibnamefont{Ni}},
  \bibinfo{author}{\bibfnamefont{H.}~\bibnamefont{Wang}},
  \bibinfo{author}{\bibfnamefont{J.}~\bibnamefont{Wu}},
  \bibinfo{author}{\bibfnamefont{Z.}~\bibnamefont{Fei}},
  \bibinfo{author}{\bibfnamefont{M.}~\bibnamefont{Goldflam}},
  \bibinfo{author}{\bibfnamefont{F.}~\bibnamefont{Keilmann}},
  \bibinfo{author}{\bibfnamefont{B.}~\bibnamefont{{\"O}zyilmaz}},
  \bibinfo{author}{\bibfnamefont{A.~C.} \bibnamefont{Neto}},
  \bibinfo{author}{\bibfnamefont{X.}~\bibnamefont{Xie}},
  \bibinfo{author}{\bibfnamefont{M.}~\bibnamefont{Fogler}},
  \bibnamefont{et~al.}, \bibinfo{journal}{Nature Materials}
  \textbf{\bibinfo{volume}{14}}, \bibinfo{pages}{1217} (\bibinfo{year}{2015}).

\bibitem[{\citenamefont{Chen et~al.}(2019{\natexlab{c}})\citenamefont{Chen,
  Sun, Wang, Gu, Xu, Wu, and Gao}}]{Chen_Science2019}
\bibinfo{author}{\bibfnamefont{W.}~\bibnamefont{Chen}},
  \bibinfo{author}{\bibfnamefont{Z.}~\bibnamefont{Sun}},
  \bibinfo{author}{\bibfnamefont{Z.}~\bibnamefont{Wang}},
  \bibinfo{author}{\bibfnamefont{L.}~\bibnamefont{Gu}},
  \bibinfo{author}{\bibfnamefont{X.}~\bibnamefont{Xu}},
  \bibinfo{author}{\bibfnamefont{S.}~\bibnamefont{Wu}}, \bibnamefont{and}
  \bibinfo{author}{\bibfnamefont{C.}~\bibnamefont{Gao}},
  \bibinfo{journal}{Science} \textbf{\bibinfo{volume}{366}},
  \bibinfo{pages}{983} (\bibinfo{year}{2019}{\natexlab{c}}).

\bibitem[{\citenamefont{Sivadas et~al.}(2018)\citenamefont{Sivadas, Okamoto,
  Xu, Fennie, and Xiao}}]{Sivadas_NanoLett2018}
\bibinfo{author}{\bibfnamefont{N.}~\bibnamefont{Sivadas}},
  \bibinfo{author}{\bibfnamefont{S.}~\bibnamefont{Okamoto}},
  \bibinfo{author}{\bibfnamefont{X.}~\bibnamefont{Xu}},
  \bibinfo{author}{\bibfnamefont{C.~J.} \bibnamefont{Fennie}},
  \bibnamefont{and} \bibinfo{author}{\bibfnamefont{D.}~\bibnamefont{Xiao}},
  \bibinfo{journal}{Nano Letters} \textbf{\bibinfo{volume}{18}},
  \bibinfo{pages}{7658} (\bibinfo{year}{2018}).

\bibitem[{\citenamefont{Jiang et~al.}(2019)\citenamefont{Jiang, Wang, Chen,
  Zhong, Yuan, Lu, and Ji}}]{Jiang_PRB2019}
\bibinfo{author}{\bibfnamefont{P.}~\bibnamefont{Jiang}},
  \bibinfo{author}{\bibfnamefont{C.}~\bibnamefont{Wang}},
  \bibinfo{author}{\bibfnamefont{D.}~\bibnamefont{Chen}},
  \bibinfo{author}{\bibfnamefont{Z.}~\bibnamefont{Zhong}},
  \bibinfo{author}{\bibfnamefont{Z.}~\bibnamefont{Yuan}},
  \bibinfo{author}{\bibfnamefont{Z.-Y.} \bibnamefont{Lu}}, \bibnamefont{and}
  \bibinfo{author}{\bibfnamefont{W.}~\bibnamefont{Ji}}, \bibinfo{journal}{Phys.
  Rev. B} \textbf{\bibinfo{volume}{99}}, \bibinfo{pages}{144401}
  (\bibinfo{year}{2019}).

\bibitem[{\citenamefont{Du et~al.}(2016)\citenamefont{Du, Wang, Liu, Hu, Utama,
  Gan, Xiong, and Kloc}}]{du2016weak}
\bibinfo{author}{\bibfnamefont{K.-z.} \bibnamefont{Du}},
  \bibinfo{author}{\bibfnamefont{X.-z.} \bibnamefont{Wang}},
  \bibinfo{author}{\bibfnamefont{Y.}~\bibnamefont{Liu}},
  \bibinfo{author}{\bibfnamefont{P.}~\bibnamefont{Hu}},
  \bibinfo{author}{\bibfnamefont{M.~I.~B.} \bibnamefont{Utama}},
  \bibinfo{author}{\bibfnamefont{C.~K.} \bibnamefont{Gan}},
  \bibinfo{author}{\bibfnamefont{Q.}~\bibnamefont{Xiong}}, \bibnamefont{and}
  \bibinfo{author}{\bibfnamefont{C.}~\bibnamefont{Kloc}}, \bibinfo{journal}{ACS
  Nano} \textbf{\bibinfo{volume}{10}}, \bibinfo{pages}{1738}
  (\bibinfo{year}{2016}).

\bibitem[{\citenamefont{Klein et~al.}(2018)\citenamefont{Klein, MacNeill, Lado,
  Soriano, Navarro-Moratalla, Watanabe, Taniguchi, Manni, Canfield,
  Fern{\'a}ndez-Rossier et~al.}}]{klein2018probing}
\bibinfo{author}{\bibfnamefont{D.~R.} \bibnamefont{Klein}},
  \bibinfo{author}{\bibfnamefont{D.}~\bibnamefont{MacNeill}},
  \bibinfo{author}{\bibfnamefont{J.~L.} \bibnamefont{Lado}},
  \bibinfo{author}{\bibfnamefont{D.}~\bibnamefont{Soriano}},
  \bibinfo{author}{\bibfnamefont{E.}~\bibnamefont{Navarro-Moratalla}},
  \bibinfo{author}{\bibfnamefont{K.}~\bibnamefont{Watanabe}},
  \bibinfo{author}{\bibfnamefont{T.}~\bibnamefont{Taniguchi}},
  \bibinfo{author}{\bibfnamefont{S.}~\bibnamefont{Manni}},
  \bibinfo{author}{\bibfnamefont{P.}~\bibnamefont{Canfield}},
  \bibinfo{author}{\bibfnamefont{J.}~\bibnamefont{Fern{\'a}ndez-Rossier}},
  \bibnamefont{et~al.}, \bibinfo{journal}{Science}
  \textbf{\bibinfo{volume}{360}}, \bibinfo{pages}{1218} (\bibinfo{year}{2018}).

\bibitem[{\citenamefont{Tong et~al.}(2018)\citenamefont{Tong, Liu, Xiao, and
  Yao}}]{tong2018skyrmions}
\bibinfo{author}{\bibfnamefont{Q.}~\bibnamefont{Tong}},
  \bibinfo{author}{\bibfnamefont{F.}~\bibnamefont{Liu}},
  \bibinfo{author}{\bibfnamefont{J.}~\bibnamefont{Xiao}}, \bibnamefont{and}
  \bibinfo{author}{\bibfnamefont{W.}~\bibnamefont{Yao}}, \bibinfo{journal}{Nano
  Letters} \textbf{\bibinfo{volume}{18}}, \bibinfo{pages}{7194}
  (\bibinfo{year}{2018}).

\bibitem[{\citenamefont{Roessler et~al.}(2006)\citenamefont{Roessler, Bogdanov,
  and Pfleiderer}}]{roessler2006spontaneous}
\bibinfo{author}{\bibfnamefont{U.~K.} \bibnamefont{Roessler}},
  \bibinfo{author}{\bibfnamefont{A.}~\bibnamefont{Bogdanov}}, \bibnamefont{and}
  \bibinfo{author}{\bibfnamefont{C.}~\bibnamefont{Pfleiderer}},
  \bibinfo{journal}{Nature} \textbf{\bibinfo{volume}{442}},
  \bibinfo{pages}{797} (\bibinfo{year}{2006}).

\bibitem[{\citenamefont{M{\"u}hlbauer et~al.}(2009)\citenamefont{M{\"u}hlbauer,
  Binz, Jonietz, Pfleiderer, Rosch, Neubauer, Georgii, and
  B{\"o}ni}}]{muhlbauer2009skyrmion}
\bibinfo{author}{\bibfnamefont{S.}~\bibnamefont{M{\"u}hlbauer}},
  \bibinfo{author}{\bibfnamefont{B.}~\bibnamefont{Binz}},
  \bibinfo{author}{\bibfnamefont{F.}~\bibnamefont{Jonietz}},
  \bibinfo{author}{\bibfnamefont{C.}~\bibnamefont{Pfleiderer}},
  \bibinfo{author}{\bibfnamefont{A.}~\bibnamefont{Rosch}},
  \bibinfo{author}{\bibfnamefont{A.}~\bibnamefont{Neubauer}},
  \bibinfo{author}{\bibfnamefont{R.}~\bibnamefont{Georgii}}, \bibnamefont{and}
  \bibinfo{author}{\bibfnamefont{P.}~\bibnamefont{B{\"o}ni}},
  \bibinfo{journal}{Science} \textbf{\bibinfo{volume}{323}},
  \bibinfo{pages}{915} (\bibinfo{year}{2009}).

\bibitem[{\citenamefont{Yu et~al.}(2010)\citenamefont{Yu, Onose, Kanazawa,
  Park, Han, Matsui, Nagaosa, and Tokura}}]{yu2010real}
\bibinfo{author}{\bibfnamefont{X.}~\bibnamefont{Yu}},
  \bibinfo{author}{\bibfnamefont{Y.}~\bibnamefont{Onose}},
  \bibinfo{author}{\bibfnamefont{N.}~\bibnamefont{Kanazawa}},
  \bibinfo{author}{\bibfnamefont{J.}~\bibnamefont{Park}},
  \bibinfo{author}{\bibfnamefont{J.}~\bibnamefont{Han}},
  \bibinfo{author}{\bibfnamefont{Y.}~\bibnamefont{Matsui}},
  \bibinfo{author}{\bibfnamefont{N.}~\bibnamefont{Nagaosa}}, \bibnamefont{and}
  \bibinfo{author}{\bibfnamefont{Y.}~\bibnamefont{Tokura}},
  \bibinfo{journal}{Nature} \textbf{\bibinfo{volume}{465}},
  \bibinfo{pages}{901} (\bibinfo{year}{2010}).

\bibitem[{\citenamefont{M{\"u}nzer et~al.}(2010)\citenamefont{M{\"u}nzer,
  Neubauer, Adams, M{\"u}hlbauer, Franz, Jonietz, Georgii, B{\"o}ni, Pedersen,
  Schmidt et~al.}}]{munzer2010skyrmion}
\bibinfo{author}{\bibfnamefont{W.}~\bibnamefont{M{\"u}nzer}},
  \bibinfo{author}{\bibfnamefont{A.}~\bibnamefont{Neubauer}},
  \bibinfo{author}{\bibfnamefont{T.}~\bibnamefont{Adams}},
  \bibinfo{author}{\bibfnamefont{S.}~\bibnamefont{M{\"u}hlbauer}},
  \bibinfo{author}{\bibfnamefont{C.}~\bibnamefont{Franz}},
  \bibinfo{author}{\bibfnamefont{F.}~\bibnamefont{Jonietz}},
  \bibinfo{author}{\bibfnamefont{R.}~\bibnamefont{Georgii}},
  \bibinfo{author}{\bibfnamefont{P.}~\bibnamefont{B{\"o}ni}},
  \bibinfo{author}{\bibfnamefont{B.}~\bibnamefont{Pedersen}},
  \bibinfo{author}{\bibfnamefont{M.}~\bibnamefont{Schmidt}},
  \bibnamefont{et~al.}, \bibinfo{journal}{Physical Review B}
  \textbf{\bibinfo{volume}{81}}, \bibinfo{pages}{041203}
  (\bibinfo{year}{2010}).

\bibitem[{\citenamefont{Yu et~al.}(2011)\citenamefont{Yu, Kanazawa, Onose,
  Kimoto, Zhang, Ishiwata, Matsui, and Tokura}}]{yu2011near}
\bibinfo{author}{\bibfnamefont{X.}~\bibnamefont{Yu}},
  \bibinfo{author}{\bibfnamefont{N.}~\bibnamefont{Kanazawa}},
  \bibinfo{author}{\bibfnamefont{Y.}~\bibnamefont{Onose}},
  \bibinfo{author}{\bibfnamefont{K.}~\bibnamefont{Kimoto}},
  \bibinfo{author}{\bibfnamefont{W.}~\bibnamefont{Zhang}},
  \bibinfo{author}{\bibfnamefont{S.}~\bibnamefont{Ishiwata}},
  \bibinfo{author}{\bibfnamefont{Y.}~\bibnamefont{Matsui}}, \bibnamefont{and}
  \bibinfo{author}{\bibfnamefont{Y.}~\bibnamefont{Tokura}},
  \bibinfo{journal}{Nature Materials} \textbf{\bibinfo{volume}{10}},
  \bibinfo{pages}{106} (\bibinfo{year}{2011}).

\bibitem[{\citenamefont{Hamamoto et~al.}(2015)\citenamefont{Hamamoto, Ezawa,
  and Nagaosa}}]{hamamoto2015quantized}
\bibinfo{author}{\bibfnamefont{K.}~\bibnamefont{Hamamoto}},
  \bibinfo{author}{\bibfnamefont{M.}~\bibnamefont{Ezawa}}, \bibnamefont{and}
  \bibinfo{author}{\bibfnamefont{N.}~\bibnamefont{Nagaosa}},
  \bibinfo{journal}{Physical Review B} \textbf{\bibinfo{volume}{92}},
  \bibinfo{pages}{115417} (\bibinfo{year}{2015}).

\bibitem[{\citenamefont{Shiomi et~al.}(2013)\citenamefont{Shiomi, Kanazawa,
  Shibata, Onose, and Tokura}}]{shiomi2013topological}
\bibinfo{author}{\bibfnamefont{Y.}~\bibnamefont{Shiomi}},
  \bibinfo{author}{\bibfnamefont{N.}~\bibnamefont{Kanazawa}},
  \bibinfo{author}{\bibfnamefont{K.}~\bibnamefont{Shibata}},
  \bibinfo{author}{\bibfnamefont{Y.}~\bibnamefont{Onose}}, \bibnamefont{and}
  \bibinfo{author}{\bibfnamefont{Y.}~\bibnamefont{Tokura}},
  \bibinfo{journal}{Physical Review B} \textbf{\bibinfo{volume}{88}},
  \bibinfo{pages}{064409} (\bibinfo{year}{2013}).

\bibitem[{\citenamefont{Fert et~al.}(2013)\citenamefont{Fert, Cros, and
  Sampaio}}]{fert2013skyrmions}
\bibinfo{author}{\bibfnamefont{A.}~\bibnamefont{Fert}},
  \bibinfo{author}{\bibfnamefont{V.}~\bibnamefont{Cros}}, \bibnamefont{and}
  \bibinfo{author}{\bibfnamefont{J.}~\bibnamefont{Sampaio}},
  \bibinfo{journal}{Nature Nanotechnology} \textbf{\bibinfo{volume}{8}},
  \bibinfo{pages}{152} (\bibinfo{year}{2013}).

\bibitem[{\citenamefont{Banerjee et~al.}(2014)\citenamefont{Banerjee, Rowland,
  Erten, and Randeria}}]{banerjee2014enhanced}
\bibinfo{author}{\bibfnamefont{S.}~\bibnamefont{Banerjee}},
  \bibinfo{author}{\bibfnamefont{J.}~\bibnamefont{Rowland}},
  \bibinfo{author}{\bibfnamefont{O.}~\bibnamefont{Erten}}, \bibnamefont{and}
  \bibinfo{author}{\bibfnamefont{M.}~\bibnamefont{Randeria}},
  \bibinfo{journal}{Physical Review X} \textbf{\bibinfo{volume}{4}},
  \bibinfo{pages}{031045} (\bibinfo{year}{2014}).

\bibitem[{\citenamefont{Johnson et~al.}(1996)\citenamefont{Johnson, Bloemen,
  Den~Broeder, and De~Vries}}]{johnson1996magnetic}
\bibinfo{author}{\bibfnamefont{M.}~\bibnamefont{Johnson}},
  \bibinfo{author}{\bibfnamefont{P.}~\bibnamefont{Bloemen}},
  \bibinfo{author}{\bibfnamefont{F.}~\bibnamefont{Den~Broeder}},
  \bibnamefont{and} \bibinfo{author}{\bibfnamefont{J.}~\bibnamefont{De~Vries}},
  \bibinfo{journal}{Reports on Progress in Physics}
  \textbf{\bibinfo{volume}{59}}, \bibinfo{pages}{1409} (\bibinfo{year}{1996}).

\bibitem[{\citenamefont{Dubowik}(1996)}]{dubowik1996shape}
\bibinfo{author}{\bibfnamefont{J.}~\bibnamefont{Dubowik}},
  \bibinfo{journal}{Physical Review B} \textbf{\bibinfo{volume}{54}},
  \bibinfo{pages}{1088} (\bibinfo{year}{1996}).

\bibitem[{\citenamefont{Banerjee et~al.}(2013)\citenamefont{Banerjee, Erten,
  and Randeria}}]{Banerjee_NatPhys2013}
\bibinfo{author}{\bibfnamefont{S.}~\bibnamefont{Banerjee}},
  \bibinfo{author}{\bibfnamefont{O.}~\bibnamefont{Erten}}, \bibnamefont{and}
  \bibinfo{author}{\bibfnamefont{M.}~\bibnamefont{Randeria}},
  \bibinfo{journal}{Nature Physics} \textbf{\bibinfo{volume}{9}},
  \bibinfo{pages}{626} (\bibinfo{year}{2013}).

\bibitem[{\citenamefont{Levy and Fert}(1981)}]{Levy_PRB1981}
\bibinfo{author}{\bibfnamefont{P.~M.} \bibnamefont{Levy}} \bibnamefont{and}
  \bibinfo{author}{\bibfnamefont{A.}~\bibnamefont{Fert}},
  \bibinfo{journal}{Phys. Rev. B} \textbf{\bibinfo{volume}{23}},
  \bibinfo{pages}{4667} (\bibinfo{year}{1981}).

\bibitem[{\citenamefont{Fert and Levy}(1980)}]{Fert_PRL1980}
\bibinfo{author}{\bibfnamefont{A.}~\bibnamefont{Fert}} \bibnamefont{and}
  \bibinfo{author}{\bibfnamefont{P.~M.} \bibnamefont{Levy}},
  \bibinfo{journal}{Phys. Rev. Lett.} \textbf{\bibinfo{volume}{44}},
  \bibinfo{pages}{1538} (\bibinfo{year}{1980}).

\bibitem[{\citenamefont{{Gilbert}}(2004)}]{1353448}
\bibinfo{author}{\bibfnamefont{T.~L.} \bibnamefont{{Gilbert}}},
  \bibinfo{journal}{IEEE Transactions on Magnetics}
  \textbf{\bibinfo{volume}{40}}, \bibinfo{pages}{3443} (\bibinfo{year}{2004}).
  
  \bibitem[{\citenamefont{Hejazi et~al.}(2020)\citenamefont{Hejazi,
  Luo, and Balents}}]{Hejazi_arXiv2020}
\bibinfo{author}{\bibfnamefont{K.}~\bibnamefont{Hejazi}},
  \bibinfo{author}{\bibfnamefont{Z.-X.}~\bibnamefont{Luo}},
  \bibnamefont{and} \bibinfo{author}{\bibfnamefont{L.}~\bibnamefont{Balents}},
  \bibinfo{journal}{arXiv:2009.00860}, (\bibinfo{year}{2020}).

\bibitem[{\citenamefont{d'Aquino et~al.}(2005)\citenamefont{d'Aquino,
  Serpico, Miano, Mayergoyz, and Bertotti}}]{d2005numerical}
\bibinfo{author}{\bibfnamefont{M.}~\bibnamefont{d'Aquino}},
  \bibinfo{author}{\bibfnamefont{C.}~\bibnamefont{Serpico}},
  \bibinfo{author}{\bibfnamefont{G.}~\bibnamefont{Miano}},
  \bibinfo{author}{\bibfnamefont{I.}~\bibnamefont{Mayergoyz}},
  \bibnamefont{and} \bibinfo{author}{\bibfnamefont{G.}~\bibnamefont{Bertotti}},
  \bibinfo{journal}{Journal of applied physics} \textbf{\bibinfo{volume}{97}},
  \bibinfo{pages}{10E319} (\bibinfo{year}{2005}).

\bibitem[{\citenamefont{Lin et~al.}(2015)\citenamefont{Lin, Saxena, and
  Batista}}]{lin2015skyrmion}
\bibinfo{author}{\bibfnamefont{S.-Z.} \bibnamefont{Lin}},
  \bibinfo{author}{\bibfnamefont{A.}~\bibnamefont{Saxena}}, \bibnamefont{and}
  \bibinfo{author}{\bibfnamefont{C.~D.} \bibnamefont{Batista}},
  \bibinfo{journal}{Physical Review B} \textbf{\bibinfo{volume}{91}},
  \bibinfo{pages}{224407} (\bibinfo{year}{2015}).

\end{thebibliography}
\end{document}